# Field-like spin orbit torque in ultra-thin polycrystalline FeMn films

Yumeng Yang[1,2], Yanjun Xu[1], Xiaoshan Zhang[1], Ying Wang[1], Shufeng Zhang[3], Run-Wei Li[4], Meysam Sharifzadeh Mirshekarloo[2], Kui Yao[2], and Yihong Wu[1,*]

[1]Department of Electrical and Computer Engineering, National University of Singapore, 4 Engineering Drive 3, Singapore 117583, Singapore

[2]Institute of Materials Research and Engineering, A*STAR (Agency for Science, Technology and Research), 08-03, 2 Fusionopolis Way, Innovis, 138634, Singapore

[3]Department of Physics, University of Arizona, Tucson, Arizona 85721, USA

[4]Key Laboratory of Magnetic Materials and Devices, Ningbo Institute of Materials Technology and Engineering, Chinese Academy of Sciences, Ningbo 315201, People's Republic of China

Field-like spin orbit torque in FeMn/Pt bilayers with ultra-thin polycrystalline FeMn has been characterized through planar Hall effect measurements. A large effective field of $2.05 \times 10^{-5}$ - $2.44 \times 10^{-5}$ Oe/(A/cm$^2$) is obtained for FeMn in the thickness range of 2 nm - 5 nm. The experimental observations can be reasonably accounted for by using a macro-spin model under the assumption that the FeMn layer is composed of two spin sublattices with unequal magnetizations. The large effective field corroborates the spin Hall origin of the effective field considering the much smaller uncompensated net moments in FeMn as compared to NiFe. The effective absorption of spin current by FeMn is further confirmed by the fact that spin current generated by Pt in NiFe/FeMn/Pt trilayers can only travel through the FeMn layer with a thickness of 1 nm – 4 nm. By quantifying the field-like effective field induced in NiFe, a spin diffusion length of 2 nm is estimated in FeMn, in consistence with values reported in literature by ferromagnetic resonance and spin-pumping experiments.

## I. INTRODUCTION

Spin-orbit torque (SOT) effect, arising from non-equilibrium spin density induced by either local or non-local strong spin-orbit interaction, has been demonstrated as a promising technique to control magnetization of ferromagnet (FM)[1-5]. Although the spin-orbit (SO) coupling induced spin polarization of electrons has been studied extensively in semiconductors, the investigations of SO induced non-equilibrium spin density in ferromagnets and the resultant SOT on local magnetization have only been reported recently. Manchon and Zhang[6] predicted theoretically that, in the presence of a Rashba spin-orbit coupling, the SOT is able to switch the magnetization of a single magnetic two-dimensional electron gas at a current density of about $10^4$–$10^6$ A/cm$^2$. This value is lower than or comparable to the critical current density of typical spin-transfer torque (STT) devices. The first experimental observation of SOT was reported by Chernyshov et al.[1] for $Ga_{0.94}Mn_{0.06}As$ dilute magnetic semiconductor (DMS) with a Curie temperature of 80 K. The $Ga_{1-x}Mn_xAs$ layer grown epitaxially on GaAs (001) substrate is compressively strained, which results in a Dresselhaus-type spin-orbit interaction that is linear in momentum. When a charge current passes through the DMS layer below its Curie temperature, the resultant SOT was able to switch the magnetization with the assistance of an external field and crystalline anisotropy. The lack of bulk inversion asymmetry (BIA) in transition metal FM has prompted researchers to explore the SOT effect in FM heterostructures with structure inversion asymmetry (SIA). Miron et al.[2] reported the first observation of a current-induced SOT in a thin Co layer sandwiched by a Pt and an $AlO_x$ layer. Due to the asymmetric interfaces with Pt and $AlO_x$, electrons in the Co layer experience a large Rashba effect, leading to sizable current-induced SOT. The Pt layer is crucial because otherwise the Rashba effect due to SIA alone would be too weak to cause any observable effect in the Co layer. At the same time, the presence of Pt also gives rise to a complex scenario about SOT in FM/heavy metal (HM) bilayers. In this case, in addition to the Rashba SOT, spin current diffused from the Pt layer due to spin Hall effect (SHE) also exerts a torque on the FM layer through transferring the spin angular momentum to the local magnetization[4]. To differentiate it from the Rashba SOT, it is also called SHE-SOT. Although the exact mechanism still remains debatable, both types of torques are generally present in the FM/HM bilayers. The former is field-like, while the latter is of anti-damping nature similar to STT. Mathematically, the two types of torques can be modeled by $\vec{T}_{FL} = \tau_{FL}\vec{m} \times (\vec{j} \times \vec{n})$ (filed-like) and $\vec{T}_{DL} = \tau_{DL}\vec{m} \times [\vec{m} \times (\vec{j} \times \vec{n})]$ (anti-damping like), respectively, where $\vec{m}$ is the magnetization direction, $\vec{j}$ is the in-plane current density, $\vec{n}$ is the interface normal, and $\tau_{FL}$ and $\tau_{DL}$ are the magnitudes of the field-like and anti-damping like torques[7-9]. Following the first report of Miron *et al.*[2], the SOT has been reported in several FM/HM bilayers with FMs such as CoFeB[5,7-10], Fe[11], NiFe[12], *etc.* and HMs such as Pt and Ta. An average effective field strength of around $4\times10^{-6}$ Oe/(A/cm$^2$) has been obtained, except for the Pd/Co multilayer system[13] which was reported to have a very large $H_{eff}/j$ value in the range of $10^{-5}$ Oe/(A/cm$^2$). A higher effective to current ratio is desirable for device applications because it will lead to a smaller critical current that is required for magnetization reversal. The critical current density for Rashba-type SOT is given by[6] $j_{critical} = \frac{\hbar e H_A M_s}{2\alpha_R m P}$, where $H_A$ is the uniaxial anisotropy field, $M_s$ the saturation magnetization, $\alpha_R$ the Rashba constant, $P$ the electron spin polarization, $m$ the electron mass, $e$ the electron charge and $\hbar$ the Planck constant. On the other hand, the anti-damping like effective field $H_{DL}$ induced by adjacent HM layer to current density ratio can be expressed as[4] $H_{DL}/j_c = \frac{\hbar}{2e}\frac{\theta_{SH}}{M_s t_{FM}}$ where $\theta_{SH}$ is the spin Hall angle of HM, $t_{FM}$ the thickness of FMs, $j_c$ the charge current. More recent studies[14,15] suggest that the spin Hall originated field-like effective field in FM/HM bilayers can also be parameterized using the same equation by replacing $\theta_{SH}$ with an effective spin Hall angle $\theta_{FL}$, *i.e.*, $H_{FL}/j_c = \frac{\hbar}{2e}\frac{\theta_{FL}}{M_s t_{FM}}$. Regardless of the role of the two types of SOT, these results suggest that FMs with low $M_s$ are desirable for investigating and exploiting the SOT effect.

Of our particular interests are antiferromagnets with small net moments due to uncompensated spins, which can potentially lead to large SOT effect in AFM/HM bilayers. In addition, AFMs are also promising for future spintronics applications due to their negligible stray field, large anisotropy and fast spin dynamics, all of which can potentially lead to AFM-based spintronic devices with improved downscaling capability, thermal stability and speed, as compared to their FM counterparts[16,17].

Unlike FM, studies on the interactions between non-equilibrium spins or spin current with AFM are quite limited. It has been predicted theoretically that spin-transfer torque (STT) can act on AFM, causing reorientation of its spin configuration, domain wall motion and stable oscillation or precession of the Neel vector[18-21]. Several follow-up experiments on exchange-biased spin-valves[22-25] have shown that current induced STT is able to affect the exchange bias at the FM/AFM interface, indirectly suggesting the presence of STT effect in AFM. More recently, spin pumping and spin torque ferromagnetic resonance (ST-FMR) measurements on FM/AFM/HM trilayers demonstrated that spin-current can travel across both NiO and IrMn at a reasonably large distance and high efficiency[26-30]. Although spin fluctuation is believed to play an important role in the spin current transport in the AFM, the real mechanism remains not well-understood at present. In addition to NiO and IrMn which have been shown to be an efficient "channel" for spin-current transport, it would be of equal interest to know if there is any AFM which shows just the opposite behavior, *i.e.*, functioning as an efficient absorber for the spin current, and if so, whether the absorbed spin current can exert a torque on the magnetization of the AFM. If such kind of AFM or phenomenon indeed exists, can we quantify the torque or effective field generated in the AFM experimentally? The answer to these questions will help to determine the potential role of AFM in future spintronic devices other than its existing role as merely a pinning layer for FM. In this regard, in this work, we investigate the spin-current induced effects in FeMn/Pt bilayers. We choose to focus on FeMn because it is the "softest" among the Mn-based AFM that have been studied for exchange bias applications; therefore, in case if there is any SOT effect in the bilayers, it can be detected easily through planar Hall effect (PHE) measurement. Recent studies have also shown that the spin Hall angle of FeMn is the smallest among PtMn, IrMn, PdMn, and FeMn[31,32]. This will facilitate the study of spin current transport across FeMn/Pt interface because the role of FeMn as a spin current generator can be neglected.

In order to investigate the SOT effect in FeMn/Pt bilayers, we fabricated a series of FeMn/Pt bilayers with different FeMn thicknesses and then characterized them through PHE measurements. Clear FM-like PHE signals were observed in FeMn/Pt bilayers with the FeMn thicknesses ranging from 2 nm to 5 nm. Magnetometry measurements of coupon films suggest that the FM-like behavior originates from canting of spin sublattices in the FeMn layer. Using the second order PHE measurement method[10,12], a field-like effective field to current ratio in the range of $2.05\times10^{-5}$ - $2.44\times10^{-5}$ Oe/(A/cm$^2$) was extracted, which is nearly two orders of magnitude larger than the typical value of $4.01\times10^{-7}$ Oe/(A/cm$^2$) for NiFe/Pt bilayers. The significantly large effective field value is understood as a result of much smaller net moments from canting of the uncompensated spins in the AFM as compared to its FM counterpart. Further investigations on NiFe/FeMn/Pt trilayers using the same PHE measurements confirm that the spin current generated by Pt is largely absorbed by FeMn and it can only travel through FeMn with a thickness of 1 nm - 4 nm. A spin diffusion length of around 2 nm in FeMn is obtained by quantifying the field-like effective field induced in NiFe, which is comparable to the ST-FMR[33] and spin pumping[32] measurements. Our results suggest that in ultra-thin polycrystalline AFMs, due to the relatively small exchange field between spin sublattices, the spin current can interact with AFM, causing reorientation of the spin sublattices, in a similar way as it does with the FM.

The remainder of this paper is organized as follows. Sec. II describes the experimental details. Sec. III A presents the structural and magnetic properties of the as-deposited FeMn film. Sec. III B discusses the magnetoresistance (MR) of NiFe/FeMn/Pt trilayer Hall bars. In Sec. III C and D, we present and discuss the electrical measurement results of FeMn/Pt bilayers. The electrical measurement results of NiFe/FeMn/Pt trilayers are presented and discussed in Sec. III E, followed by conclusions in Section IV.

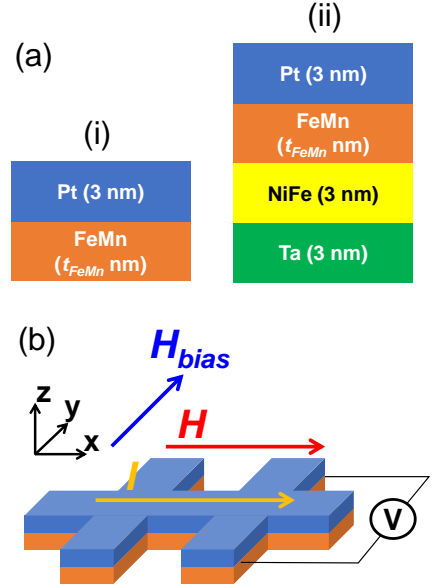

FIG. 1. (a) Schematic of FeMn/Pt bilayer (i) and NiFe/FeMn/Pt trilayers (ii) samples; (b) Schematic of the second order PHE measurement setup with a transverse bias field ($H_{bias}$).

## II. EXPERIMENTAL DETAILS

As illustrated in Figs. 1(a), two series of samples in the form of Hall bars (Fig. 1(b)) were prepared on $SiO_2$(300 nm)/Si substrates with the following configurations: (i) Si/SiO$_2$/FeMn($t_{FeMn}$)/Pt(3) and (ii) Si/SiO$_2$/Ta(3)/NiFe(3)/FeMn($t_{FeMn}$)/Pt(3) (number in the parentheses indicates the thickness in nm). The thickness ($t_{FeMn}$) of FeMn was varied in the range of 0 - 15 nm to investigate its effect on transport properties. Throughout this manuscript, we adopt the convention that multilayer always start from the substrate side, *e.g.*, FeMn/Pt refers to Si/SiO$_2$/FeMn/Pt. The Hall bars, with a central area of 2.3 mm × 0.2 mm and transverse electrodes of 0.1 mm × 1 mm, were fabricated using combined techniques of photolithography and sputtering deposition. The former was performed using a Microtech laserwriter and the latter was carried out using a DC magnetron sputter with a base and process pressure of $3 \times 10^{-8}$

Torr and process 3 mTorr, respectively. During the deposition of the trilayers, an in-plane bias field of ~500 Oe was applied along the long axis of the Hall bar to induce an in-plane easy axis in NiFe. The resistivity of individual layers was extracted from the overall resistivity of bilayers with thicknesses in the same range of those for transport measurements but with different thickness combinations, and the obtained resistivity values are: $\rho_{Ta}$ = 159 μΩ·cm, $\rho_{NiFe}$ = 79 μΩ·cm, $\rho_{FeMn}$ = 166 μΩ·cm, and $\rho_{Pt}$ = 32 μΩ·cm.

All electrical measurements were carried out at room temperature using the Keithley 6221 current source and 2182A nanovolt meter. The PHE measurements were performed by supplying a DC bias current ($I$) to the Hall bar and measuring the Hall voltage ($V_{xy}$) while sweeping an external field ($H$) in x-axis direction (see schematic in Fig. 1(b)). Second order PHE measurements were carried out to quantify the spin current induced effective field in both FeMn/Pt bilayers and NiFe/FeMn/Pt trilayers[10,12]. In this method, a set of second order PHE voltages, defined as $\Delta V_{xy}(H_{bias}) = V_{xy}(+I, +H_{bias}, H) + V_{xy}(-I, -H_{bias}, H)$, are obtained from the algebraic sum of the first order Hall voltages measured at a positive (+$I$) and negative bias (-$I$) current, respectively, at three different transverse bias fields in y-axis direction: –$H_{bias}$, 0 and $H_{bias}$. Here, $I$ is the current applied, $H$ is the external field in x-axis direction, and $V_{xy}$ is the first order Hall voltage. Under the small perturbation assumption, *i.e.*, both the current induced field ($H_{FL}$) and applied transverse bias field ($H_{bias}$) are much smaller than the external field ($H$), the change in in-plane magnetization direction is proportional to $(H_I + H_{bias})/H_{eff}$, where $H_I$ is the sum of $H_{FL}$ and Oersted field ($H_{Oe}$), and $H_{eff}$ is the sum of $H$ and anisotropy field ($H_A$). The linear dependence of second order PHE voltage on the algebraic sum of $H_I$ and $H_{bias}$ allows one to determine the effective field by varying $H_{bias}$ as both fields play an equivalent role in determining the magnetization direction. After some algebra, it is derived that $\frac{\Delta V_{xy}(0)}{\Delta V_{xy}(H_{bias}) - \Delta V_{xy}(-H_{bias})} = \frac{H_{FL} + H_{Oe}}{2H_{bias}}$. By linearly fitting $\Delta V_{xy}(0)$ against $[\Delta V_{xy}(H_{bias}) - \Delta V_{xy}(-H_{bias})]$, the ratio of $(H_{FL}+H_{Oe})$ to $2H_{bias}$ can be determined from the slope of the curve. After subtraction of $H_{Oe}$ from $H_I$, the current induced $H_{FL}$ at a specific bias current can thus be obtained. Although the second order PHE method was initially developed for quantifying the effective field in NiFe/Pt bilayers, as we will discuss later, it can also be applied to FeMn/Pt bilayers by dividing the FeMn into two spin sublattices with unequal magnetizations. The same procedure can also be used to determine the effective field in NiFe in NiFe/FeMn/Pt trilayers as in this case the PHE signal is mainly from the NiFe layer and the signal from FeMn can be neglected.

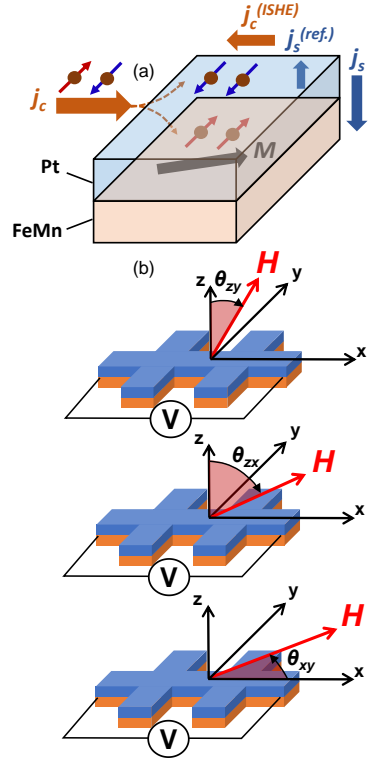

FIG. 2. (a) Schematic of SMR generation mechanism in FeMn/Pt bilayers; (b) Schematic of ADMR measurements with a constant rotating field $H$ in $zy$, $zx$, and $xy$ planes, respectively.

To further confirm the SOT effect in FeMn/Pt bialyers, spin Hall magnetoresistance (SMR) measurements were performed on these bilayers with different FeMn thicknesses. It has been reported in the FM/HM cases[34,35], SMR has the same origin with the damping-like effective field $H_{DL}$. As shown in the schematic of Fig. 2(a), when a charge current $j_c$ flows in x-direction, a spin current $j_s$ is generated from the Pt layer through SHE. The spin current follows in z-direction with the spin polarization $\vec{\sigma}$ in y-direction. When the spin current reaches the FeMn/Pt interface, depending on the angle between the magnetization $\vec{M}$ of FeMn and $\vec{\sigma}$, a certain portion of the spin current is reflected back into Pt with the remaining traveling across the interface and absorbed by FeMn. The reflection is maximum when $\vec{M} \parallel \vec{\sigma}$ and minimum when $\vec{M} \perp \vec{\sigma}$. The reflected spin current $j_s^{(ref.)}$ is converted to a charge current $j_c^{(ISHE)}$ in Pt through the inverse spin Hall effect (ISHE) which flows in the opposite direction of the original current $j_c$. As a consequence, the longitudinal resistance of Pt in x-direction is modulated by the direction of $\vec{M}$, leading to the appearance of SMR given by $R_{xx} = R_0 - \Delta R (\vec{m} \cdot \vec{\sigma})^2$, where $R_{xx}$ is the longitudinal resistance, $\vec{m}$ the unit vector of magnetization, $R_0$ the isotropic longitudinal resistance, and $\Delta R$ the SMR induced resistance change[36]. As illustrated in Fig. 2(b), the SMR can be readily obtained by measuring $R_{xx}$ under a rotating magnetic field in different coordinate planes, or angle-dependent magnetoresistance (ADMR) measurements. If the applied field $H$ is sufficiently large to saturate the magnetization, the SMR ratio can be calculated from the relation $\Delta R / R_{xx} = \left( R_{xx}^z - R_{xx}^y \right) / R_{xx}^y$, where $R_{xx}^z$ and $R_{xx}^y$ are the longitudinal resistance $R_{xx}$ obtained with $H$ applied in z- and y-direction, respectively. The value of SMR and SOT effective field are closely related to each other in the way that SMR (SOT) is minimum (maximum) when $\vec{M} \perp \vec{\sigma}$ and *vice versa* when $\vec{M} \parallel \vec{\sigma}$. The main difference is that the reflected spin current is converted to SMR through ISHE whereas the FeMn absorbed spin current is converted to SOT effective field through the magnetic moment in FeMn. Therefore, the observation of clear SMR can further confirm the SOT effect observed in the FeMn/Pt layer (a more quantitative discussion will be presented in Sec. III C).

In addition to Hall bars, coupon films have also been prepared for X-ray diffraction and magnetic measurements. The XRD measurements were performed on D8-Advance Bruker system with Cu K$_\alpha$ radiation. Magnetic measurements were carried out using a Quantum Design vibrating sample magnetometer (VSM) with the samples cut into a size of 4 mm × 5 mm. The resolution of the system is better than $6\times10^{-7}$ emu.

## III. RESULTS AND DISCUSSION
### A. Structural and magnetic properties of FeMn

Fig. 3 shows the XRD patterns of coupon films with different structures: (A) Si/SiO$_2$/Ta(3)/NiFe(3)/FeMn(15)/Ta(3), (B) Si/SiO$_2$/Ta(3)/FeMn(15)/Ta(3), (C) Si/SiO$_2$/FeMn(15)/Ta(3), and (D) Si/SiO$_2$/Ta(3)/NiFe(3)/Ta(3). The Ta capping layer is used to prevent the samples from oxidization. In order to obtain a certain level of signal strength, the thickness of FeMn was intentionally made thicker than those of the samples for electrical transport measurements. As can be seen from the figure, all the samples with a FeMn layer, namely, A, B, and C, exhibit a peak at 43.5°, corresponding to the (111) peak of FeMn. This indicates that the FeMn layer is well textured in [111] direction. The bottom Ta layer enhances the adhesion to the substrate, but it has negligible effect on the texture of FeMn as shown by the subtle difference between the peak intensities of XRD pattern B and C. Therefore, for electrical measurements, the Ta seed layer can be removed in order to avoid the formation of dead layer at the Ta/FeMn interface and also to eliminate any current induced effect from Ta. On the other hand, the insertion of a thin NiFe underlayer significantly enhances the [111] texture of FeMn, as can be seen from the significantly larger peak intensity of A as compared to B and D.

Magnetic measurements were performed on two series of coupon films: (i) a single layer of FeMn(3) covered by different capping layers: Pt(3), Ta(3), and Au(3); and (ii) a single layer of FeMn($t_{FeMn}$) with $t_{FeMn}$ = 1 nm - 15 nm capped by a Pt(3) layer. Fig. 4(a) shows the magnetization versus field (M-H) loops for the first set of samples after subtracting the diamagnetic signal from the substrate. All the samples exhibit FM-like M-H curves with a negligible hysteresis but a large saturation field around

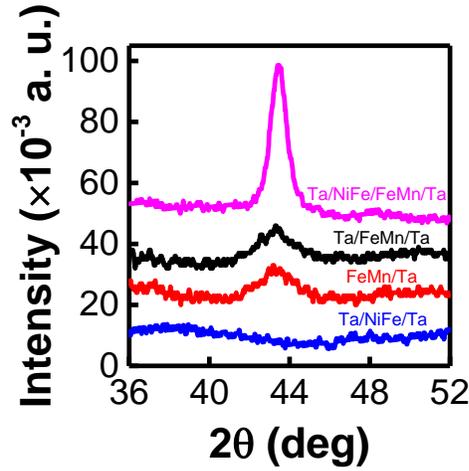

FIG. 3. XRD patterns for Ta(3)/NiFe(3)/FeMn(15)/Ta(3), Ta(3)/FeMn(15)/Ta(3), FeMn(15)/Ta(3) and Ta(3)/NiFe(3)/Ta(3) coupon films. Curves are vertically shifted for clarity.

20 kOe. The samples capped with Pt and Au show similar M-H loops and saturation magnetization, whereas the sample capped by Ta exhibit an apparently different behavior: both the saturation field and magnetization are much smaller than those of the other two samples. As shown in the inset of Fig. 4(a), the saturation magnetization $M_s$ (averaged over the field range from 20 kOe to 30 kOe) of Pt capped sample is slightly higher than that of the Au capped sample, and both are almost double of that of the Ta capped sample. This is consistent with earlier reports that (1) Pt interfacial layer can be easily magnetized through proximity effect when contacting with a FM[37,38], but the same type of effect is weak in Au[39] and (2) Ta can create magnetic dead layer in the adjacent FM[40]. Similar proximity effect has been observed at FeMn/Pt interfaces in previous studies on exchange bias[41,42]. Obviously the proximity effect induced moment alone is unable to account for the large saturation moment shown in Fig. 4(a). In order to better understand the origin of the observed net moment, VSM measurements were performed on the second series of samples with varying FeMn thicknesses but a fixed Pt capping layer. Fig. 4(b) shows the M-H loops of FeMn($t_{FeMn}$)/Pt(3) with $t_{FeMn}$ = 2 nm, 3 nm, 5 nm, 8 nm, and 15 nm, respectively. Although the shape of the M-H loops looks quite similar among these samples, the saturation magnetization decreases quickly with increasing $t_{FeMn}$ and it drops to almost zero at $t_{FeMn}$ = 8 nm (see Fig. 4(c)). This suggests that the observed saturation magnetizations in thin FeMn are mainly due to canting of spin sublattices subject to a large external field. Canting at a moderate field is only possible when the thickness is small due to the reduced sublattice exchange field at small thickness. With the increase of thickness, a bulk-like AFM order will eventually be full established; when this happens it would be difficult to cause any canting of the spin sublattices at a moderate field, leading to a vanishing saturation magnetization in the FeMn/Pt bilayer. Any residual saturation moment observed in samples with thick FeMn must come from both the proximity induced moment in Pt and the uncompensated spins from the interfacial layer of FeMn. These net moments are expected to decrease quickly from the interface. However, when $t_{FeMn}$ is below $t_0$ (the critical thickness for establishing a rigid AFM order at room temperature), as depicted in Fig. 4(d), the interaction between Pt and FeMn will lead to formation of two spin sublattices with unequal magnetizations. Although the net uncompensated moment is expected to decrease from the interface, for the sake of simplicity, we will assume that it is uniform throughout the FeMn when it is thin.

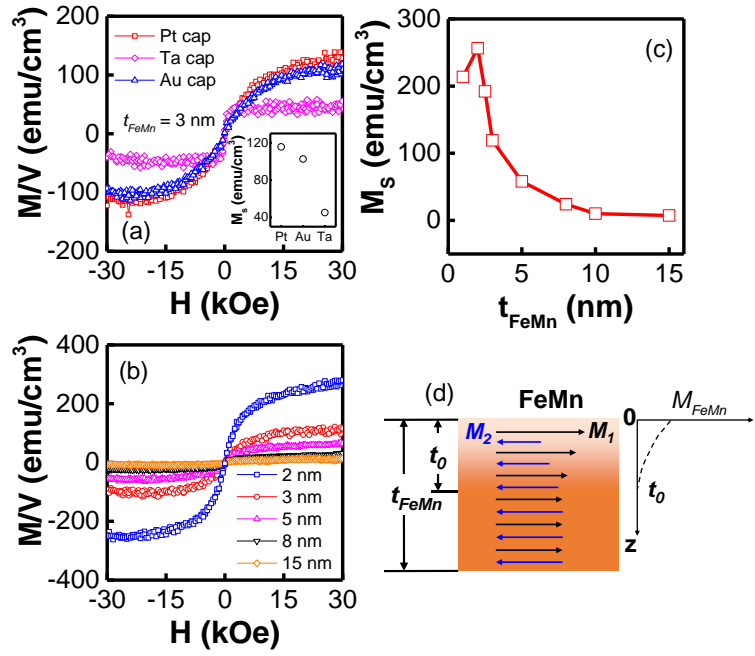

FIG. 4. (a) M-H loops for FeMn(3)/Pt(3), FeMn(3)/Ta(3), FeMn(3)/Au(3), respectively; (b) M-H loops for FeMn($t_{FeMn}$)/Pt with $t_{FeMn}$ = 2 nm, 3 nm, 5 nm, 8 nm, and 15 nm; (c) FeMn thickness dependence of $M_s$ of FeMn($t_{FeMn}$)/Pt (3) bialyers; (d) Illustration of spin sublattices with unequal magnetizations in FeMn near the FeMn/Pt interface. Inset of (a): $M_s$ of bilayers with different capping layer.

**B. Magnetoresistance of NiFe/FeMn/Pt trilayers**

To further correlate the magnetic property of FeMn with the M-H loops in Fig. 4, magnetoresistance (MR) measurements were performed on NiFe(3)/FeMn($t_{FeMn}$)/Pt trilayer Hall bars with $t_{FeMn}$ varying from 0 to 15 nm. Figs. 5(a) and 5(b) show the MR curves for samples with $t_{FeMn}$ in the range of 0 – 5 nm and 8 – 15 nm, respectively. Since the MR from NiFe is significantly larger than that of FeMn, we can safely assume that the MR is dominated by the signal from NiFe for all the samples, regardless of the FeMn thickness. Shown in Fig. 5(c) are the coercivity of NiFe ($H_c$) and exchange bias field ($H_{eb}$) at the NiFe/FeMn interface extracted from the MR curves in Figs. 5(a) and 5(b). As can be seen from the results, the effect of FeMn on NiFe depends strongly on its thickness. For $t_{FeMn} < 2$ nm, there is neither $H_c$ enhancement of NiFe nor observable $H_{eb}$ at the NiFe/FeMn interface. This indicates that in this thickness region the blocking temperature ($T_B$) and possibly Neel temperature ($T_N$) of the magnetic grains are below room temperature (RT). In other words, the spin sublattices within each grain are weakly coupled and the entire film behaves more or less like a superpara-antiferromagnet. At $t_{FeMn}$ of 3 nm - 5 nm, an increased $H_c$ (around 8 – 270 Oe) and a small $H_{eb}$ (around 1 – 3 Oe) were observed, suggesting the formation of AFM order ($T_N > T_B > $ RT) as the thicknesses increases. In this case, the exchange coupling between the spin sublattices should have already been established in most of the grains, though its strength as well as the anisotropy remains small and varies among the different grains. Therefore, in this thickness region, the FeMn layer may be treated as an AFM with a finite distribution of exchange coupling strength and anisotropy, with both having a small magnitude. As a consequence, the AFM sublattices can be canted by an external magnetic field with a moderate strength, as shown in Figs. 4(a) and 4(b). The onset of a clear exchange bias, with the $H_{eb}$ (~450 Oe) comparable to typical values reported in literature[43], was observed for samples with $t_{FeMn}$ > 8 nm. In this thickness range, the variation in exchange coupling among the grains may be ignored, and the entire film can be treated as an AFM with a uniform exchange coupling strength, but having a finite distribution of anisotropy. As reproduced in the inset of Fig. 5(c), the observed thickness dependence of the AFM order in our FeMn film is consistent with the previous theoretical calculation[44] of the thickness dependence of $T_B$. It should be noted that the critical thickness for onset of clear exchange bias coincides with the thickness above which the saturation magnetization drops to a minimum in Figs. 4(b) and 4(c). This further affirms our explanation that the large saturation moments observed in thin FeMn are due to canting of the spin sublattices. As will be presented shortly, the current-induced PHE signal also vanishes as the thickness of FeMn exceeds the critical thickness in both bilayer and trilayer samples. Therefore, we focus the discussion hereafter mainly on ultra-thin FeMn films (1 nm – 5 nm). Although the FeMn layers in this thickness range are not normal AFM in the strict sense, the improved response of AFM spins to external field provides a convenient way to study the interaction of AFM with spin current.

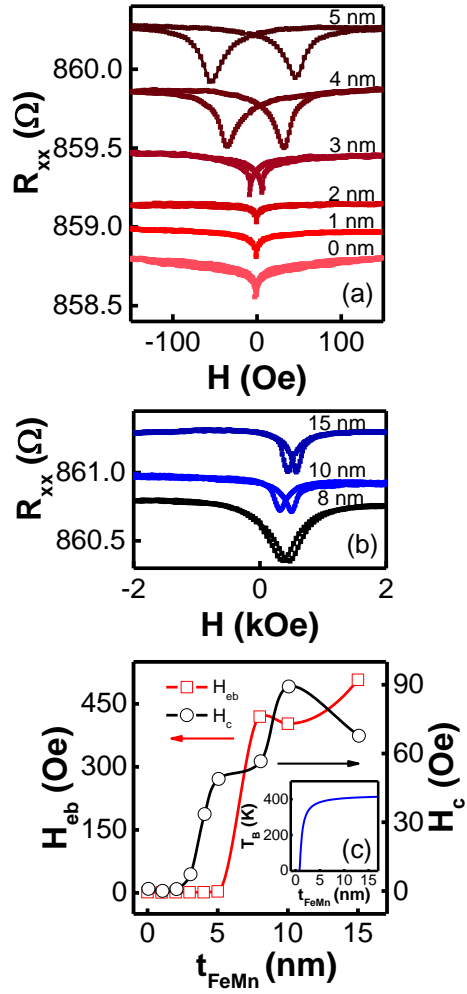

FIG. 5. (a) MR curves for NiFe(3)/FeMn($t_{FeMn}$)/Pt trilayers with $t_{FeMn}$ = 0 – 5 nm; (b) MR curves for NiFe(3)/FeMn($t_{FeMn}$)/Pt trilayers with $t$ = 8 nm – 15 nm; (c) Dependence of $H_c$ and $H_{eb}$ on $t_{FeMn}$ extracted from (a) and (b). Inset of (c): $t_{FeMn}$-dependence of $T_B$ (reproduced from Ref. 44).

### C. PHE measurements of FeMn/Pt bilayers

We now turn to the PHE measurement results of FeMn($t_{FeMn}$)/Pt(3) bilayer samples. The measurement geometry is shown in Fig. 6(a). Shown in Fig. 6(b) are the planar Hall resistance ($\Delta R_{xy}$) versus field ($H$) curves obtained at different bias currents ($I$), for the $t_{FeMn}$ = 3 nm sample. Here, the Hall resistance is given by $\Delta R_{xy} = [V_{xy}(+I, H) + V_{xy}(-I, H)]/2I$, which represents the change in Hall resistance caused by the current-induced effective field. As can be seen from Fig. 6(b), the overall shape of the PHE curves resembles that of a typical FM. The Hall signal is weak at low bias current and increases prominently with increasing the bias current. Moreover, the peak position of PHE shifts to larger field values as the bias current increases. Since the AFM consists of grains with randomly distributed in-plane anisotropy axes, the PHE signal can be understood as resulting from two competing fields, *i.e.*, the externally applied field in *x*-direction and current-induced effective field in *y*-direction, acting on the spin sublattices of FeMn. The increase of PHE signal amplitude and shift of the peak position can be understood as being caused by the increase of $H_I$ when the current increases. The role of $H_I$ is confirmed by the observation that the PHE signal vanishes when the field is swept in *y*-direction, as shown in Fig. 6(c) for a bias current of 5 mA. To further demonstrate that $H_I$ indeed originates from the spin Hall effect, we fabricated a Si/SiO$_2$/FeMn(3)/Ta(3) control sample. Fig. 6(d) shows the comparison of the PHE curves at 5 mA for both FeMn(3)/Ta(3) and FeMn(3)/Pt(3) samples. A similar FM-like PHE signal is observed in FeMn/Ta except that the magnitude is much smaller and its polarity is opposite to that of FeMn/Pt. The latter implies that the sign of $H_I$ in FeMn/Ta is opposite to that of FeMn/Pt, which is consistent with the opposite sign of $\theta_{SH}$ for Pt and Ta. It can also be inferred from the results that Joule heating is not the major cause for the observation, because otherwise one would expect a PHE with same polarity in both FeMn/Pt and FeMn/Ta as the temperature gradient is not likely to change direction upon changing the top layer as both Pt and Ta have a lower resistivity as compared to FeMn. The bias current dependence of PHE for samples with different FeMn thickness is similar to the one shown in Fig. 6(b) except that its magnitude decreases with increasing the FeMn thickness. Fig. 6(e) shows the FeMn thickness dependence of PHE voltage. To have a meaningful comparison, instead of showing the nominal Hall resistance by dividing the Hall voltage by the total current, we show the Hall voltage scaled by the currents in both the FeMn ($I_{FeMn}$) and Pt ($I_{Pt}$) layer. This makes sense because the PHE signal mainly comes from the FeMn layer but its amplitude is determined by the current-induced field ($H_I$) from the Pt layer. $I_{FeMn}$ and $I_{Pt}$ were calculated using three-dimensional (3D) finite element analysis by using the experimentally derived resistivity values for different layers given in Section II. To shorten the simulation time, the Hall bar sample was scaled down to a strip with a length of 2 μm, a width of 0.2 μm and the thicknesses of each layer remained the same as the actual samples. As can be seen from Fig. 6(e), the PHE signal decreases with increasing the FeMn thickness, and it becomes vanishingly small at thicknesses above 8 nm (not shown here). This is in good agreement with the results of both the VSM and MR measurements, as discussed above. In other words, the PHE signal observed in FeMn/Pt bilayers are caused by the current-induced canting of spin sublattices with unequal magnetizations. The signal gradually decreases to zero as the AFM hardens with increasing the thickness.

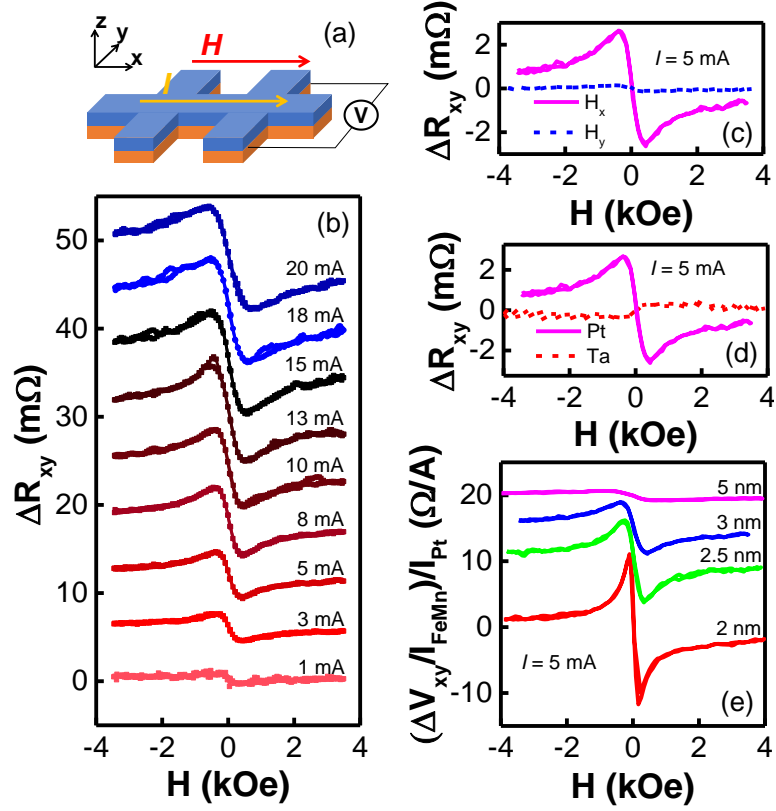

FIG. 6. (a) Schematic of PHE measurement at different bias currents; (b) PHE curves for FeMn(3)/Pt(3) at different bias currents; (c) PHE curves for FeMn(3)/Pt(3) obtained at 5 mA with field swept in *x*- and *y*-direction, respectively; (d) A comparison of PHE curves at 5 mA for FeMn(3)/Ta(3) (dashed line) and FeMn(3)/Pt(3) (solid line) with the field applied in *x*-direction; (e) Normalized PHE curves for samples with different FeMn thickness from 2 nm – 5 nm. Note that curves in (b) and (e) are vertically shifted for clarity.

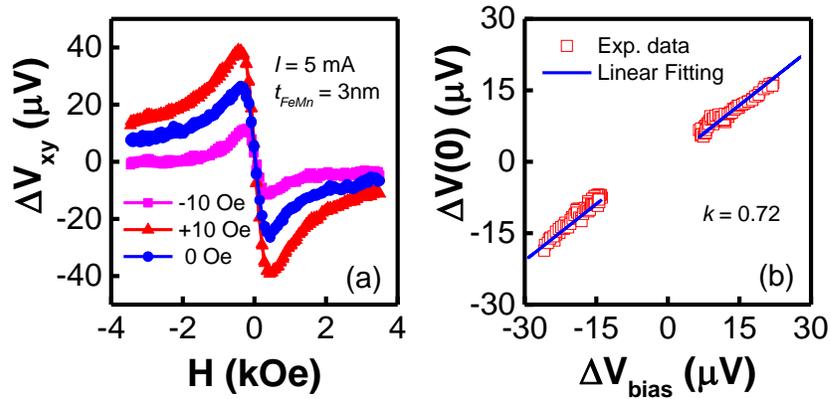

FIG. 7. (a) PHE curves for the FeMn(3)/Pt(3) bilayer measured at 5 mA with different transverse bias field (0 Oe, +10 Oe and -10 Oe); (b) Linear fitting of $\Delta V_{xy}$ (0) against $\Delta V_{bias} = [\Delta V_{xy} (H_{bias} = 10$ Oe$) - \Delta V_{xy} (H_{bias} = -10$ Oe$)]$ to determine the ratio of the current-induced $H_I$ to $2H_{bias}$.

In order to quantify the strength of $H_I$, we carried out the second order PHE measurements as described in Sec. II. Fig. 7(a) shows an example of one set of PHE curves with $H_{bias}$ = 0 Oe, +10 Oe and – 10 Oe, respectively, at a bias current of 5 mA for the FeMn(3)/Pt(3) sample. As can be seen, the PHE signal magnitude changes with the total field in y-direction including both $H_I$ and $H_{bias}$. The increase of PHE at $H_{bias}$ = +10 Oe indicates that $H_I$ is in positive y-direction. Fig. 7(b) shows the linear fitting of $\Delta V_{xy}(0)$ against $\Delta V_{bias} = [\Delta V_{xy}(+10$ Oe$) - \Delta V_{xy}(-10$ Oe$)]$ using the data in Fig. 7(a). For a better linear approximation, the data at low fields were excluded and only the data at fields above ±1 kOe were used for the fitting[10]. $H_I$ can be calculated from the slope $k$ by using the relation $H_I = 2kH_{bias}$. The offset between the fitting lines at positive and negative region is understood to be caused by either $H_{DL}$ or the thermal effect[10,12]. The small amplitude of the offset confirms again that the contributions from both effects are small in the PHE signals obtained from the FeMn/Pt bilayers. The same experiments have been repeated for FeMn/Pt bilayers with different FeMn thickness ($t_{FeMn}$ = 2 nm – 5 nm), and the results are shown in Fig. 8(a). As can be seen, the $H_I$ in all samples scales almost linearly with the bias current. After subtracting the Oersted field ($H_{Oe}$), the effective-field ($H_{FL}$) normalized to the current density in Pt is shown in Fig. 8(b). The Oersted field in the FeMn layer is calculated using 3D finite element analysis on scaled down strips with a dimension of 20 μm × 2 μm. The calculated Oersted field ($H_{Oe}$) (also normalized to the current density in Pt) in the order of $1 \times 10^{-7}$ Oe/(A/cm$^2$) is almost independent of the FeMn thickness and is much smaller than the measured $H_I$ for all samples. As shown in in Fig. 8(b), the $H_{FL}/j_{Pt}$ ratio (open square) is in the range of $2.05 \times 10^{-5}$ - $2.44 \times 10^{-5}$ Oe/(A/cm$^2$) for FeMn/Pt bilayers; this is nearly two orders of magnitude larger than that of the NiFe/Pt control sample ($4.01 \times 10^{-7}$ Oe/(A/cm$^2$)). Although the physical origin of the field-like effective field in FM/HM bilayers is still debatable, recent studies suggest that it can be written in the following form by taking into account the spin Hall current from the HM layer only[45,46]:

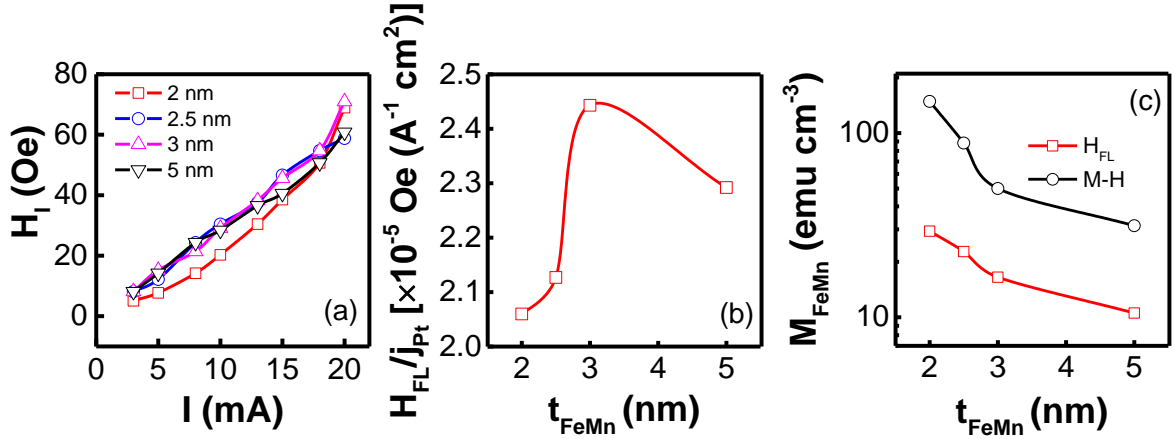

FIG. 8. (a) Extracted $H_I$ for FeMn($t_{FeMn}$)/Pt(3) bilayers with $t_{FeMn}$ = 2 nm – 5 nm; (b) $H_{FL}/j_{Pt}$ (open square) as a function of $t_{FeMn}$ after subtracting the Oersted field; (c) $M_{FeMn}$ calculated from $H_{FL}$ using Eq. (1) (open square) and $M_{FeMn}$ extracted from the M-H loops at 4 kOe (open circle). Note that the data in (c) is plotted in log scale for clarity.

$$H_{FL}/j_c = \frac{\hbar}{2e}\frac{\theta_{SH}}{M_s t_{FM}}(1-\frac{1}{\cosh(d/\lambda_{HM})}) \times \frac{g_i}{(1+g_r)^2 + g_i^2} \qquad (1)$$

where $g_r = \text{Re}[G_{MIX}]\rho\lambda_{HM}\coth(d/\lambda_{HM})$, $g_i = \text{Im}[G_{MIX}]\rho\lambda_{HM}\coth(d/\lambda_{HM})$ with $G_{MIX}$ the spin mixing conductance of FM/HM interface, $\rho$ the resistivity of HM, and $\lambda_{HM}$ the spin diffusion length in HM. The spin Hall origin of the field-like effective field is supported by several experimental studies[7,11,12,14], especially when the FM layer is thick, based on the observation that the field directions are opposite to each other in Pt and Ta based FM/HM bilayers with a same FM. Following this scenario, the large effective field obtained in this study can be readily understood by substituting the relevant parameters into Eq. (1). These include the moment per unit area in NiFe ($M_s t_{NiFe}$) and FeMn ($M_{FeMn} t_{FeMn}$) and spin mixing conductance ($G_{MIX}$) at the NiFe/Pt and FeMn/Pt interfaces. If we assume a same $G_{MIX}$ for the two types of interfaces and use the known $M_s$ of NiFe of 800 emu/cm$^3$, the resultant net magnetization of FeMn, $M_{FeMn}$, is in the range of 10.5 – 29.3 emu/cm$^3$ with a thickness of 2 nm – 5 nm, as shown in Fig. 8(c) (open square). Also shown in Fig. 8(c) (open circle) is the average magnetization extracted from the M-H curves shown in Fig. 4(b) at an applied field of 4 kOe (note: we use the magnetization at 4 kOe instead of the saturation magnetization because the maximum applied field in electrical measurements was 4 kOe). As can be seen from the figure, although the net magnetization from M-H loops is around 5 times larger than that calculated from the $H_{FL}$, both show very similar trend as long as FeMn thickness dependence is concerned. The difference in absolute values is understandable because in electrical measurements the magnetic moment that affects $H_{FL}$ is mainly concentrated at the FeMn/Pt interface, whereas the VSM measurement detects the moment of the entire film. These results suggest that the small net moment is the determining factor that gives the large effective field to current ratio as compared to NiFe.

As shown in Fig. 8(b), the electrically derived $H_{FL}/j_{Pt}$ ratio (open square) increases sharply with FeMn thickness below 3 nm and then decreases slowly as $t_{FeMn}$ increases further. This is in sharp contrast with the monotonically decreasing dependence of $H_{FL}$ on FM thickness ($t_{FM}$) in typical FM/HM heterostuctures[12,47]. The latter is due to the fact when $t_{FM}$ increases, the product of $t_{FM}$ and $M_{FM}$ increases accordingly, leading to a $1/t_{FM}$ dependence of $H_{FL}$. However, in the case of FeMn/Pt bilayers, the net magnetization $M_{FeMn}$ decreases with $t_{FeMn}$ (> 2 nm), as confirmed by the VSM measurement results shown in Fig. 8(c). This naturally leads to a peak in the curve in Fig. 8(b). The peak position of $H_{FL}$ agrees well with the region where $H_C$ is enhanced but clear exchange bias has yet to be established (see Fig. 5(c)). This suggests that the enhancement of $H_{FL}$ occurs in the region that AFM order is just about to form and their spin sublattices can still be canted easily by either an external or effective field. We noticed that in early theoretical work on spin torque in AFM, $H_{FL}$ is treated as negligibly small[48,49]. This is valid for rigid AFM systems. It should be pointed out that our results presented in Figs. 6 - 8 do not contradict these reports because the $H_{FL}$ indeed vanishes when the FeMn thickness is above 8 nm. At such thickness, a rigid AFM order is formed and any $H_{FL}$ on the spin sublattices should have been cancelled out.

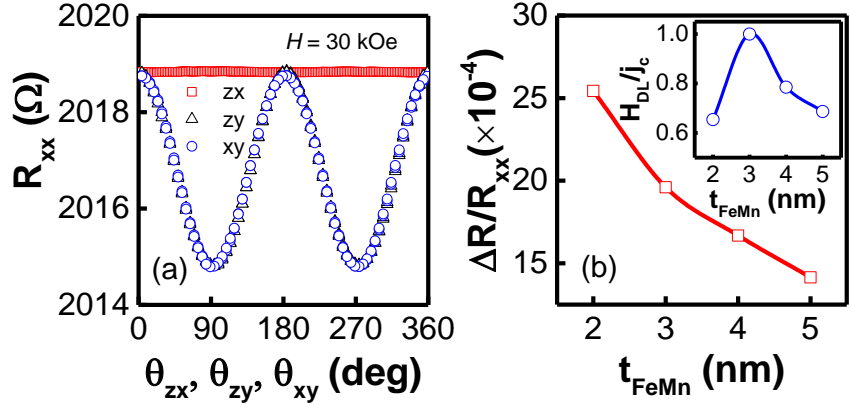

FIG. 9. (a) ADMR results at 30 kOe for FeMn(3)/Pt(3) bilayer; (b) Thickness dependence of SMR ratio $\Delta R/R_{xx}$ with $t_{FeMn}$ = 2 nm – 5 nm. Inset of (b): Normalized thickness dependence of damping like effective field calculated from Eq. (4).

To further confirm the SOT effect in FeMn/Pt and verify the non-monotonic thickness dependence of the effective field, we performed the ADMR measurements in the bilayer samples with $t_{FeMn}$ = 2 nm – 5 nm using the schematic shown in Fig. 2(b). Fig. 9(a) shows the ADMR results for a FeMn(3)/Pt(3) bilayer measured at a constant field of 30 kOe while rotating the sample in *xy, zx, zy* planes, respectively. The almost overlapping between $\theta_{zy}$- and $\theta_{xy}$- dependence of MR indicates that the conventional anisotropic MR in FeMn/Pt is negligibly small and the MR measured is dominated by SMR. The SMR ratio on the order of $10^{-3}$ is comparable to that in NiFe/Pt reported earlier[50], and much larger than that in YIG/Pt system[36]. Fig. 9(b) shows the SMR ratio as a function of FeMn thickness in the range $t_{FeMn}$ = 2 nm – 5 nm, which decreases monotonically as the FeMn thickness increases, suggesting the decrease of spin current entering the FeMn layer. To have a quantitative correlation of this thickness dependence to that of $H_{DL}$, one needs to look into their expressions, respectively. Firstly, the SMR ratio can be expressed as[34,46]:

$$\frac{\Delta R}{R_{xx}} = \theta_{SH}^2 \frac{\lambda_{Pt}}{d_{Pt}} \frac{\tanh(d_{Pt}/2\lambda_{Pt})}{(1+\xi)}(1-\frac{1}{\cosh(d_{Pt}/\lambda_{Pt})}) \times \frac{g_r(1+g_r)+g_i^2}{(1+g_r)^2+g_i^2} \quad (2)$$

where $\xi = \rho_{Pt} t_{FeMn} / \rho_{FeMn} d_{Pt}$ is introduced to take into account the current shunting effect by FeMn, and $\rho_{Pt}$ ($\rho_{FeMn}$) and $d_{Pt}$ ($t_{FeMn}$) are the resistivity and thickness of Pt (FeMn), respectively. On the other hand, the damping-like effective field $H_{DL}$ can be written as[45,46]:

$$H_{DL} / j_c = \frac{\hbar}{2e} \frac{\theta_{SH}}{M_s t_{FeMn}} (1 - \frac{1}{\cosh(d_{Pt} / \lambda_{Pt})}) \times \frac{g_r(1+g_r) + g_i^2}{(1+g_r)^2 + g_i^2} \quad (3)$$

The combination of Eq. (2) and (3) gives:

$$H_{DL} / j_c = \frac{\hbar}{2e} \frac{1}{\theta_{SH} M_s t_{FeMn}} \frac{d_{Pt}}{\lambda_{Pt}} \frac{1}{\tanh(d_{Pt} / 2\lambda_{Pt})} \frac{\Delta R}{R_{xx}} \quad (4)$$

Note that we have set $\xi = 0$ in Eq. (4) since the current shunting effect taken into account in the calculation of SMR has nothing to do with the reflection/transmission of spin current at the FeMn/Pt interface, or in any case it is much smaller than unity due to the large difference in resistivity between Pt and FeMn. In this way, the thickness dependence of $H_{DL}/j_c$ can be readily calculated from Eq. (4) by using the thickness dependence of SMR obtained in Fig. 9(b). The inset of Fig. 9(b) shows the normalized FeMn thickness dependence of damping-like effective field calculated from Eq. (4). Note that ideally, we should use the moment of FeMn at the interface only for $M_{FeMn} t_{FeMn}$. However, as it is difficult to extract the interface moment independently, we used the volumetric $M_{FeMn}$ instead, which was obtained by the VSM measurement in Fig. 8(c). Although it is not exactly the same, the thickness dependence of $H_{DL}$ is indeed similar to the FeMn thickness dependence of $H_{FL}$ presented in Fig. 8(b). Therefore, from the results obtained by second order PHE and ADMR measurements, we demonstrated clearly the existence of SOT effect in FeMn/Pt and the non-monotonic dependence of the SOT effective field on FeMn thickness.

**D. Macro-spin model of the FeMn layer**

In order to have a more quantitative understanding of the M-H loops in Fig. 4(b) and PHE curves in Fig. 6(b) for the FeMn/Pt bilayers, we have simulated both curves using the macro-spin model.

Although the spin state of bulk FeMn can take either a collinear or non-collinear configuration[51-54], the spin configuration in an ultrathin film may differ from that of the bulk, especially when it interacts with FM or HM like Pt. In the case of FeMn/FM bilayer, it has been observed experimentally that the spin axis of FeMn is aligned to that of the FM layer from the interface[55-57]. In the case of FeMn/Pt bilayers, the situation can be more complicated due to the strong spin-orbit interaction of Pt. Determination of the exact spin configuration is beyond the scope of this work which certainly deserves further investigations. However, in order to simplify the problem yet without compromising the underlying physics, we treat ultrathin FeMn layer as being consisting of two collinear spin sublattices with unequal saturation magnetizations $M_s$. As we will show in this section, the good agreement between experimental and simulation results supports the collinear model. Under this assumption, the M-H loops and PHE curves of FeMn/Pt bilayers shown previously can be simulated through energy minimization. Based on the coordinate notation in Fig. 10(a), the free energy density $E$ of a specific grain in the FeMn layer can be written as[58]

$$E = J|\vec{M}_1||\vec{M}_2|\cos(\theta_1 - \theta_2) - H[|\vec{M}_1|\cos(\varphi - \theta_1) + |\vec{M}_2|\cos(\varphi - \theta_2)] + K_u(\sin^2\theta_1 + \sin^2\theta_2) \quad (5)$$

where $J$ is the sublattice exchange coupling constant, $|\vec{M}_1|$ and $|\vec{M}_2|$ are the magnitude of $\vec{M}_1$ and $\vec{M}_2$, respectively, $\theta_1$ and $\theta_2$ are the angles of $\vec{M}_1$ and $\vec{M}_2$ with respect to y-direction, respectively, $\varphi$ is the angle between y-direction and H, and $K_u$ is the uniaxial anisotropy constant. Eq. (5) can be solved numerically to find the steady-state values for $\theta_1$ and $\theta_2$, which in turn can be used to calculate the M-H curve. To facilitate the discussion, we introduce the following parameters: $N = |\vec{M}_1|/|\vec{M}_2|$, $H_A = K_u/|\vec{M}_2|$ and $H_{ex} = J|\vec{M}_2|$. Note that Eq. (5) applies to a single grain with a specific anisotropy axis and exchange coupling strength. Considering the polycrystalline nature of the sample, ideally one should simulate the average M-H curve by taking into account the finite distribution of anisotropy axes and exchange field. However, it is found that the calculated curve with a fixed anisotropy axis at 0° is very similar to the one that is obtained by assuming that the anisotropy axes is distributed from 0° – 90° at a step of 10° and then taking an average of the calculated curves at different angles. This is due to the fact that $K_u$ in ultra-thin FeMn is small, and its effect on steady-state magnetization direction is overtaken by the current-induced effective field. Therefore, for simplicity, in the subsequent simulations we assumed that the uniaxial anisotropy is along y-axis for all the grains. A log-normal distribution was adopted to account for the exchange field ($H_{ex}$) distribution: $f(H_{ex}) = \frac{1}{H_{ex}\sigma\sqrt{2\pi}}\exp[-\frac{(\ln H_{ex} - \mu)^2}{2\sigma^2}]$, with $\mu = \log(5000)$, and $\sigma = 0.5$ when $H_{ex}$ is in unit of Oe. This is justifiable because the grain size of sputtered polycrystalline films typically follows the lognormal distribution[59] and the AFM order is found to enhance with the increase of grain size[60]. The average M-H curve was obtained by assuming $H_{ex}$ in the range of 1 kOe – 19 kOe with a step of 2 kOe. As can be seen from Fig. 10(b), a reasonably good agreement is obtained between the simulated (solid line) and experimental M-H curves for the $t_{FeMn} = 3$ nm sample by assuming $N = 1.2$, $H_A = 50$ Oe, and $M_s = 115.83$ emu/cm$^3$. Next, we proceed to account for the spin current in the sample by introducing in Eq. (5) an additional Zeeman energy terms arising from $H_{FL}$, i.e. $-H_{FL}(|\vec{M}_1|\cos\theta_1 + |\vec{M}_2|\cos\theta_2)$. Similarly, $\theta_1$ and $\theta_2$ are determined numerically at different $H_{FL}$ values which in turn are used to calculate the normalized PHE signal at different $H$: $PHE = (|\vec{M}_1|\sin 2\theta_1 + |\vec{M}_2|\sin 2\theta_2)/(|\vec{M}_1| + |\vec{M}_2|)$. Fig. 10(c) compares the simulated curves at different $H_{FL}$ values with the field in $x$-direction. The simulated curve resembles typical PHE curve for a FM and the peak position increases with increasing $H_{FL}$, both of which agree well with experimental PHE curves obtained at different bias currents. As shown in the inset of Fig. 10(c), when the field is changed to $y$-direction, a vanished PHE is obtained. Therefore, the macro-spin model is able to account for the main experimental observations in FeMn/Pt bilayers. This strongly supports our arguments that the large field-like spin orbit torque in FeMn/Pt bilayers is caused by the relatively small magnetic moment in the FeMn, and resultant SOT is able to induce canting of the spin sublattices of the AFM.

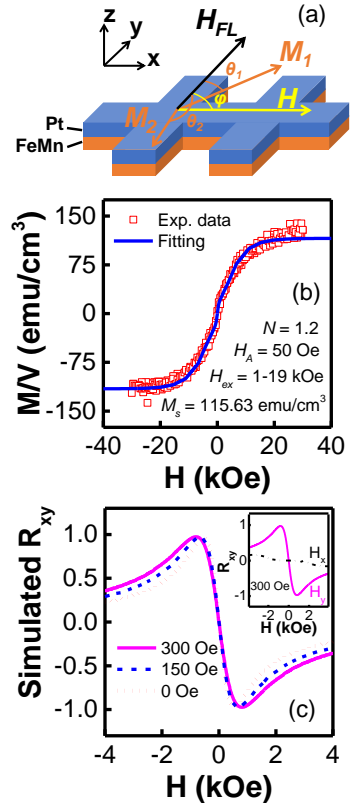

FIG. 10. (a) Illustration of the FeMn spin sublattice configuration, external field and current-induced $H_{FL}$; (b) M-H loop fitting using the macro-spin model for FeMn(3)/Pt(3); (c) Simulated PHE curves with different $H_{FL}$ values (0 Oe, 150 Oe and 300 Oe). Inset of (d): Simulated PHE curves at $H_{FL}$ = 300 Oe with the external field applied in $x$ - and $y$ - direction, respectively.

Before ending this section, we would like to comment on the validity of the macro-spin model. Although the films are polycrystalline, we argue that the macro-spin model is able to capture the essential physics of current-induced SOT in FeMn/Pt bilayers because unlike the charge current which flows in the lateral direction (*i.e.*, *x*-direction), the spin current generated from Pt flows mainly in *z*-direction (*i.e.*, in the sample normal direction). Since the FeMn thickness in the samples under investigation (2 nm – 5 nm) is comparable to the grain size, we can safely assume that the spin current is confined mostly inside a single crystal grain with negligible influence from the grain boundaries (different from the laterally flowing charge current). Therefore, as long as the polycrystalline film has a well-defined texture in the thickness direction which is the case in this study, it would appear locally as a "quasi-single crystal" to the vertically flowing spin-current. Compared to the true single crystal case, the only difference is that in the polycrystalline case, the SOT effect is further averaged over different grains due to the random distribution of crystalline anisotropy and exchange energy, which has been taken into account in the above discussion. Therefore, we believe the macro-spin model is appropriate for interpretation of the experimental results observed in this work.

### E. PHE measurements of NiFe/FeMn/Pt trilayers

To further demonstrate that the spin current generated in Pt is indeed largely absorbed by FeMn, we have fabricated NiFe(3)/FeMn($t_{FeMn}$)/Pt(3) trilayer Hall bars and studied SOT-induced magnetization rotation in NiFe. Fig. 11(a) shows the PHE curves at different bias currents ($I$) for the NiFe(3)/FeMn(3)/Pt(3) sample. Similar to the results shown in Fig. 6(b), the PHE signal increases prominently as $I$ increases, indicating the presence of a current-induced effective field $H_I$ in $y$-direction. The Hall signal is much larger than that of the FeMn/Pt bilayer in the same field range; therefore the signal from the trilayer is dominantly from the NiFe layer. The results can be qualitatively understood as follows. The spin current generated by the Pt layer travels through the FeMn spacer and induces SOT in the NiFe layer. The SOT will then cause a rotation of the NiFe magnetization, leading to the observed increase of PHE with the bias current. To have a more quantitative understanding of the current dependence of PHE signal, 3D micromagnetic modeling was performed on an NiFe element with and without a transverse field using OOMMF[61]. To shorten the computation time, in the simulation, the sample is scaled down to a strip with a dimension of 23 μm × 2 μm × 3 nm. The parameters used are: saturation magnetization $M_s = 8 \times 10^5$ A/m, exchange constant $J = 1.3 \times 10^{-11}$ J/m, damping constant $\alpha = 0.5$, anisotropy constant $K_u = 100$ J/m$^3$ and unit cell size: 10 nm × 10 nm × 3 nm. A fixed bias field in $y$-direction is used to simulate the effective field induced by the current. To account for the Hall measurement geometry, only the data at the center area of 1 μm × 2 μm representing the Hall bar cross is taken into consideration for the calculation of PHE signal. Fig. 11(b) shows the simulated PHE curves at bias fields of 0 Oe, 5 Oe and 10 Oe, respectively. Note that due to the much smaller size used in the simulation, the simulated $H_c$ is much larger than the measured value, and therefore a large transverse bias field of 10 Oe was used in the simulation accordingly. Except for the large $H_c$, the simulated curves resemble well the measured PHE curves. Fig. 11(c) shows the normalized PHE curves for samples with different FeMn thicknesses at a bias current of 10 mA. As can be seen, the signal amplitude decreases as the thickness increases, indicating the decrease of the $H_I$ at larger FeMn thickness. When the FeMn thickness exceeds 5 nm, the signal becomes vanishingly small, suggesting that the spin current cannot travel through the FeMn layer beyond this thickness.

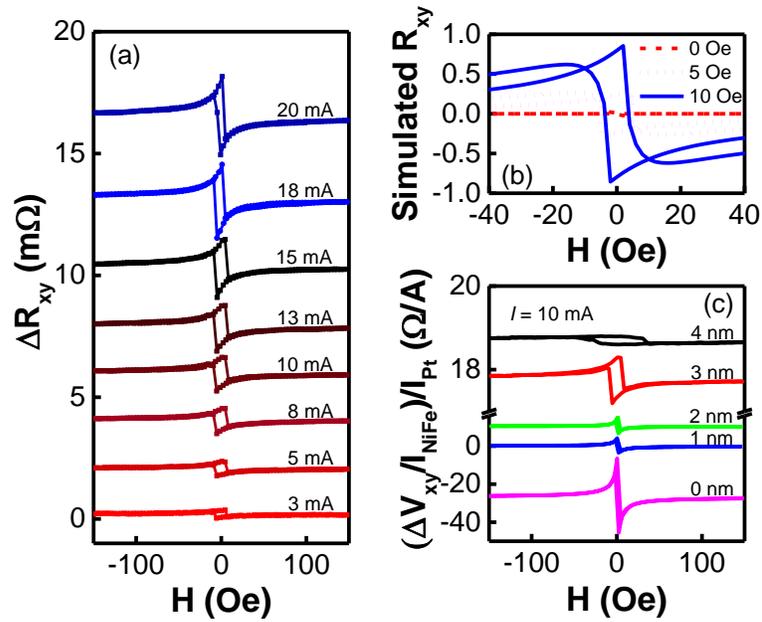

FIG. 11. (a) PHE curves at different bias currents for the NiFe(3)/FeMn(3)/Pt(3) trilayer; (b) Simulated PHE curves with 0 Oe, 5 Oe and 10 Oe bias field in *y*-direction; (c) Normalized PHE curves at 10 mA for the trilayer sample with FeMn thicknesses of 0 – 4 nm. Note that the curves in (a) and (c) are vertically shifted for clarity.

To quantity the strength of the field-like effective field in the NiFe layer, again we carried out the second order PHE measurements. Fig. 12(a) shows one set of PHE curves for NiFe(3)/FeMn(3)/Pt(3) obtained with $I = 10$ mA, and $H_{bias} = 0$ Oe, +0.6 Oe and – 0.6 Oe, respectively. The flip of curve polarity at positive and negative bias field suggests that $H_I$ is comparable to the applied bias field of 0.6 Oe. Fig. 12(b) shows the linear fitting of $\Delta V_{xy}(0)$ against $\Delta V_{bias}$ using the data in Fig. 12(a). The slope $k$ turns out to be much smaller than that obtained for the FeMn/Pt bilayers, as shown in Fig. 7(b). This in turn gives a much smaller $H_I$ for the trilayer samples with $t_{FeMn} = 0 - 4$ nm, as shown in Fig. 12(c). Similar to the case of FeMn/Pt bilayers, $H_I$ for all samples scales almost linearly with the bias current. The $t_{FeMn} = 0$ sample corresponds to a Ta(3)/NiFe(3)/Pt(3) trilayer. The obtained $H_I$ value of 0.52 Oe at a bias current of 10 mA is comparable to reported value for similar structure[12]. The $H_I$ value drops sharply with the insertion of a 1 nm FeMn, and decreases further as the FeMn thickness increases. To quantify the current contribution directly from the Pt layer, we have to subtract from $H_I$ two other contributions, *i.e.*, $H_{Oe}$ in the NiFe layer and $H_{FL}$ from the Ta seed layer. The total Oersted field in NiFe, $H_{Oe}$, is calculated using 3D finite element analysis, and the results are shown in the inset of Fig. 12(d) as a function of FeMn thickness (down triangle); it increases with FeMn thickness due to the increase of current in the FeMn layer. In order to estimate the contribution of current in the Ta layer to $H_I$, we have fabricated a NiFe(3)/Ta(3) control sample and measured the effective field using the same second order PHE measurement. The effective field to current ratio obtained is $H_{FL}(Ta)/j_{Ta} = 1.49\times10^{-7}$ Oe/(A/cm$^2$). Based on this value, we can estimate the contribution of Ta current in the trilayers with different FeMn thicknesses. The results are shown in the inset of Fig. 12(d) in upper triangles. The value of $H_{FL}(Ta)$ is almost constant due to the much larger resistivity of Ta as compared to other layers. Also shown in the inset is the FeMn thickness dependence of $H_I$. The net effective field is obtained as $H_{FL} = H_I - H_{Oe} - H_{FL}(Ta)$. As shown in Fig. 12(d), all the samples exhibit a non-zero $H_{FL}$ except for the $t_{FeMn} = 4$ nm sample in which $H_I$ and $H_{Oe}$ are comparable. As shown clearly in the inset of Fig. 12(d), the contribution of Ta layer to the effective field is negligible.

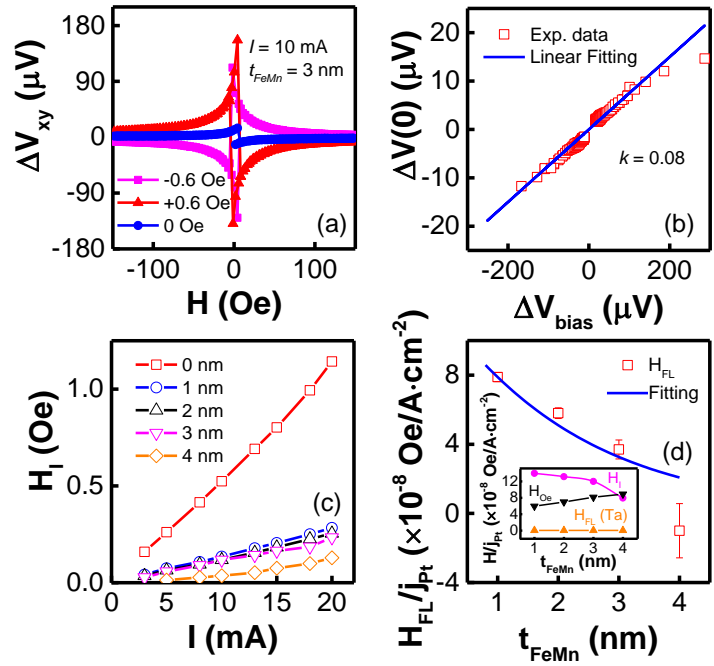

FIG. 12. (a) PHE curves for the NiFe(3)/FeMn(3)/Pt(3) trilayer measured at 10 mA with different transverse bias field (0 Oe, +0.6 Oe and -0.6 Oe); (b) Linear fitting of $\Delta V_{xy}(0)$ against $\Delta V_{bias} = [\Delta V_{xy}(H_{bias} = 0.6$ Oe$) - \Delta V_{xy}(H_{bias} = -0.6$ Oe$)]$ to determine the ratio of the current-induced $H_I$ to $2H_{bias}$; (c) Extracted $H_I$ for samples with $t_{FeMn} = 0$ nm – 4 nm; (d) Experimental values for $H_I$ (open square) and fitting using Eq. (8) (solid line). Inset of (d): FeMn thickness dependence of $H_I$ (circle), $H_{Oe}$ in NiFe (down triangle) and $H_{FL}$ from Ta (upper triangle), respectively. Note that the data in (d) are normalized to the current density in Pt.

After excluding the contribution from Ta as main source, the net $H_{FL}$ must be induced by the spin current from the Pt layer since the spin Hall angle of FeMn is very small[31,32]. Considering the fact that the Pt layer has a same thickness in all the samples, it is plausible to assume that the spin Hall angle and thickness scaling factor $[1-1/\cosh(d/\lambda_{HM})]$ of Pt are the same among the different samples. We further assume that the moment per unit area of NiFe ($M_s t_{NiFe}$) is also a constant. Therefore, the decrease in effective field in the NiFe layer can only come from two sources: (1) relaxation of spin current in FeMn and (2) reduced spin mixing conductance ($G_{MIX}$) at the FeMn/Pt and NiFe/FeMn interfaces as compared to the single NiFe/Pt interface. Earlier reports[32,33] found that spin transport in FM/normal metal (NM)/FeMn structures is mainly dependent on the FM/NM interface and the spin relaxation inside FeMn. Therefore, rather than a dramatic modification of $G_{MIX}$ at the interfaces with the presence of the FeMn layer, the absorption of spin current by FeMn is more likely the major cause for decreased spin current entering NiFe. This spin absorption explanation is also consistent with the large $H_{FL}$ observed in FeMn/Pt bilayers.

The spin current in the NiFe layer induced by Pt in the NiFe/FeMn/Pt trilayer can be modeled using the drift-diffusion approach. Due to the relatively large size of the Hall bar sample in the $xy$ plane, the spin current can be treated as non-equilibrium spins flowing in $z$-direction with polarization in $y$-direction. Therefore, the spatial distribution of spin current in NiFe/FeMn can be written as:

$$j_i(z) = -\frac{1}{2e\rho_i}\frac{\partial \Delta\mu_i(z)}{\partial z} \tag{6}$$

where $i = 1$ refers to FeMn, $i = 2$ denotes NiFe, $\Delta\mu_i$ and $j_i$ are the net spin accumulation and spin current density in layer $i$, respectively, and $\rho_i$ is resistivity of layer $i$. The spin accumulation satisfies the following diffusion equation[62]:

$$\frac{\partial^2 \Delta\mu_i(z)}{\partial z^2} = \frac{\Delta\mu_i(z)}{\lambda_i^2} \tag{7}$$

where $\lambda_i$ is the spin diffusion length of layer $i$. The general solution for $\Delta\mu_i$ is $\Delta\mu_i(z) = A_i \exp(z/\lambda_i) + B_i \exp(-z/\lambda_i)$. To obtain specific solutions, we need to set up proper boundary conditions. As discussed above, the effect of Ta layer is negligible. In order to obtain a simple analytical solution, we assume that the spin current is zero at the NiFe/Ta interface. Based on this assumption, we adopted the following boundary conditions: $j_1(0) = j_0$, $j_2(t_2) = 0$, $j_1(t_1) = j_2(t_1)$ and $\Delta\mu_1(t_1) = \Delta\mu_2(t_1)$, where $t_1$ is the thicknesses of the FeMn ($t_{FeMn}$), $t_2$ is the sum of the thickness of FeMn and NiFe layer ($t_{FeMn} + t_{NiFe}$), and $j_0$ is the spin current generated by Pt entering FeMn. Substituting the boundary conditions into Eq. (6) and (7), the spin current density at the interface entering NiFe can be derived as:

$$j(t_1)/j_0 = \frac{2\lambda_1\rho_1 A(1-B^2)}{\lambda_1\rho_1(1+A^2)(1-B^2) + \lambda_2\rho_2(1-A^2)(1+B^2)} \tag{8}$$

where $A = \exp(t_{FeMn}/\lambda_1), B = \exp(t_{NiFe}/\lambda_2)$. Comparing it with Eq. (1), we can see that the spin absorption in FeMn layer gives an additional scaling factor for spin current to be delivered to the NiFe layer. In the extreme case when $t_{NiFe}$ approaches infinite, i.e., $B \to \infty$, Eq. (8) is reduced to $j(t_1)/j_0 \approx 1/A$, if $\lambda_1\rho_1 \approx \lambda_2\rho_2$, which is the exponential decay formula used in Ref. 27, 30 and 33 to obtain the spin diffusion length in AFMs. On the other hand, if $t_1 = 0$, $j(t_1)/j_0 = 1$, which means that the spin-current generated by Pt will enter NiFe directly without absorption in the FeMn layer. In our sample, since the NiFe thickness is comparable to that of FeMn, the effect of NiFe can no longer be ignored. Note that the difference in $G_{MIX}$ of NiFe/Pt and FeMn/Pt interfaces is ignored for simplicity and we also assume that $G_{MIX}$ is independent of FeMn thicknesses. Although from the results in Fig. 9(b) it may be inferred that $G_{MIX}$ is thickness dependent (i.e. $j_0$ is dependent on $t_{FeMn}$), in the above derivation we mainly focus on the spin current decay in FeMn and consider $j_0$ as a constant. By scaling the $H_{FL}$ obtained in NiFe layer using the resistivity of the films obtained above and the spin diffusion length of NiFe ($\lambda_2 = 3$ nm)[63], as shown in Fig. 12(d), the spin diffusion length of FeMn ($\lambda_1$) is obtained as 2 nm. This value is comparable to earlier reports of 1.9 nm (Ref. 33) and 1.8 ± 0.5 nm (Ref. 32). The short spin diffusion length is consistent with the previous understanding of AFM as a good "spin sink"[64,65]. The effective absorption of spin current by FeMn is consistent with the large SOT effect observed in FeMn/Pt bilayers. Although the spin configuration of FeMn in the bilayer sample may be different from that of the trilayer sample due to the insertion of the NiFe seed layer in the latter, we foresee that the difference, if any, is only qualitative; it will not affect the results and conclusion drawn in this section in a fundamental way.

The difference in FeMn thickness dependence of $H_{FL}$ between the bilayer case (Fig. 8(b)) and trilayer (Fig. 12(d)) case can be understood as follows. As we discussed in Sec. III C (see Fig. 9(b)), although the spin current traveling across FeMn/Pt deceases almost linearly with $t_{FeMn}$, the $H_{FL}$ in FeMn/Pt bilayer is mainly determined by the thickness dependence of the magnetic moment in FeMn ($M_{FeMn}t_{FeMn}$) (see Fig. 8(c)). On the other hand, for the NiFe/FeMn/Pt trilayer case, $H_{FL}$ is for the NiFe layer (the signal from FeMn is masked out by that of NiFe due to its much smaller magnetization), and thus it is a measure of spin current that travels across the FeMn layer and eventually enters the NiFe layer. As can be seen from Eq. (8), the spin current traveling in FeMn further decays by a factor of $\frac{2\lambda_1\rho_1 A(1-B^2)}{\lambda_1\rho_1(1+A^2)(1-B^2)+\lambda_2\rho_2(1-A^2)(1+B^2)}$ upon reaching the NiFe/FeMn interface. This decay, together with the almost linear decay of SMR (see Fig. 8(b)) gives the overall decay of spin current upon reaching the NiFe/FeMn interface. This spin current is further converted to $H_{FL}$ in NiFe through the magnetic moment ($M_{NiFe}t_{NiFe}$). Since the NiFe thickness is fixed among the samples, the FeMn thickness dependence of $H_{FL}$ in NiFe of the trilayers should be the same as that of the spin current reaching the NiFe/FeMn interface. This explains why the $H_{FL}$ in NiFe decreases monotonically with the FeMn thickness, which is different from that in FeMn.

Before we conclude, it is worth pointing out that the FeMn investigated in this work has a polycrystalline structure, and due to the ultra-thin thickness, the AFM order may not be well defined as in the bulk material. We foresee this as the main challenge in investigating and exploiting SOT effect in AFM materials, *i.e.*, SOT is more prominent in ultra-thin layers, but most AFM requires a finite thickness to develop a stable AFM order at room temperature. To overcome this difficulty, it is necessary to development AFM materials which allow effective generation of non-equilibrium spins in the bulk. One of the possible candidates is AFM with bulk inversion asymmetry and strong SO interaction[48].

**IV. CONCLUSIONS**

In summary, our systematic studies revealed that spin Hall current from Pt induces SOT in the FeMn layer in FeMn/Pt bilayers, which is able to induce canting of the spin sublattices of FeMn when its thickness is below 5 nm. Based on current-dependent PHE measurements, a large field-like effective field of $2.05 \times 10^{-5}$ - $2.44 \times 10^{-5}$ Oe/(A/cm$^2$) was obtained for FeMn in the thickness range of 2 nm - 5 nm, which is attributed to the small net moment in FeMn as compared to its FM counterpart. The origin of the moment was further investigated by the magnetometry measurements, and is found to be mainly from FeMn itself arising from the canting of the uncompensated spin sublattices. The spin-canting process can be explained reasonably well based on the macro-spin model by taking into account the current-generated effective field. Further investigations on NiFe/FeMn/Pt trialyers show that spin current from Pt is strongly absorbed by the FeMn layer with a spin diffusion length of around 2 nm, which explains why the SOT effect is strong in FeMn/Pt bilayers when $t_{FeMn}$ is small and becomes negligible when $t_{FeMn} > 10$ nm. Although it remains a challenge to ensure the presence of both well-defined AFM order and large SOT in thin AFM layers, the results presented here shall stimulate further studies on spin transport in AFM materials with different types of crystalline and spin structures.

## ACKNOWLEDGMENTS

The authors wish to thank Prof. Jingshen Chen, Dr. Kaifeng Dong and Dr. Baoyu Zong from National University of Singapore, and Dr. Wendong Song of Data Storage Institute for their assistance in magnetic measurements. Y.H.W. would like to acknowledge support by the Singapore National Research Foundation, Prime Minister's Office, under its Competitive Research Programme (Grant No. NRF-CRP10-2012-03) and Ministry of Education, Singapore under its Tier 2 Grant (Grant No. MOE2013-T2-2-096). M.S.M. and K.Y. acknowledge the support of IMRE, A*STAR (Agency for Science, Technology and Research) under project IMRE/10-1C0107. R.-W L. acknowledges the support of the National Natural Foundation of China (11274321) and the State Key Project of Fundamental

Research of China (2012CB933004). S. Z. is partially supported by US-NSF-ECCS-1127751. Y.H.W. is a member of the Singapore Spintronics Consortium (SG-SPIN).

* Electronic mail: elewuyh@nus.edu.sg